# Grain boundary strain localization in CdTe solar cell revealed by Scanning 3D X-ray diffraction microscopy

A. Shukla [a]*, J. Wright [b], A. Henningsson [c], H. Stieglitz [d], E. Colegrove [e], L. Besley [a], C. Baur [a], S De Angelis [a], M. Stuckelberger [f], H.F. Poulsen [a], J.W. Andreasen [a]

[a] Technical University of Denmark, Fysikvej Building 310, Lyngby, Denmark, [b] European Synchrotron Radiation Facility, 71 Av. des Martyrs, Grenoble, France, [c] Lund University, Ole Römers väg 1, Lund, Sweden [d] Helmholtz-Zentrum Hereon, Max-Planck-Straße 1, 21502 Geesthacht, Germany, Germany, [e] National Renewable Energy Laboratory, Golden, CO 80401, United States of America, [f] Deutsches Elektronen-Synchrotron DESY, Notkestraße 85, Hamburg, Germany

**Abstract**

Cadmium Telluride (CdTe) solar cell technology is a promising candidate to help boost green energy production. However, impurities and structural defects are major barriers to improving the solar power conversion efficiency. Grain boundaries often act as aggregation sites for impurities, resulting in strain localization in areas of high diffusion. In this study, we demonstrate the use of scanning 3D X-ray diffraction microscopy to non-destructively make 3D maps of the grains – their phase, orientation, and local strain – within a CdTe solar cell absorber layer with a resolution of 100 nm. We quantify twin boundaries and suggest how they affect grain size and orientation distribution. Local strain analysis reveals that strain is primarily associated with high misorientation grain boundaries, whereas twin boundaries do not show high strain values. We also observe that high-strain grain boundaries form a continuous pathway connected to the CdS layer. Hence, this high-strain region is believed to be associated with the diffusion of sulfur from the CdS layer along grain boundaries. This hypothesis is supported by SEM-EDS and X-ray fluorescence experiments. The method and analysis demonstrated in this work can be applied to different polycrystalline materials where the characterization of grain boundary properties is essential to understand the microstructural phenomena.

**Introduction**

Cadmium Telluride (CdTe) solar cells have emerged as a promising candidate to boost green energy production. This is vital in pursuing the UN Sustainable Development Goal of providing clean and affordable energy to all by 2030. They are commercially one of the most successful solar cell technologies with an installed capacity of 50 GW. With their direct band gap suitable for absorbing incoming solar radiation, CdTe solar cells have demonstrated a high-power conversion efficiency (PCE) of above 22.3%[1] and are cost-competitive with other utility-scale solar cell technologies[1]. Despite the notable progress made, there is still significant potential for improving the conversion efficiency of thin-film CdTe solar cells. The performance remains significantly below the Shockley-Queisser theoretical limit[2], primarily owing to the prevailing voltage deficit.

This voltage deficit results from charge carrier recombination in the bulk of polycrystalline solar cells, which has been reported to increase at grain boundaries[3]. Close-space sublimation (CSS) followed by $CdCl_2$ treatment is a mature physical vapor deposition (PVD) technique used for industrial manufacturing of CdTe solar cells. Here, the growth of thin films follows a Volmer-Weber growth model[4]. As the isolated islands nucleate on the substrate surface, coarsen, and coalesce, they form grain boundaries that are under high biaxial tensile stress[5]. This stress is subsequently relieved by the diffusion of impurity atoms towards grain boundaries[6] and by deformation twinning. (Twinning is common in CdTe due to a low formation energy of stacking faults[7]). Both mechanisms are prevalent in vapor deposited CdTe thin films. Further annealing under $CdCl_2$ vapor environment increases extent of impurity diffusion and twinning [4,8].

Some of the diffused impurities which have been reported are sulfur, chlorine, copper, and sodium[9]. Grain boundaries can act as aggregation sites for these impurities because of the large surface area and, hence, higher surface energy[10]. It is energetically more favorable for impurities to move towards grain boundaries than to stay in the grain interior[11]. This aggregation leads to an increase in the number of trap states in the band gap of the solar cell near the grain boundary and hence increases the probability of charge-carrier recombination. More specifically, grain boundary-assisted diffusion of sulfur has been reported in numerous studies. Yan et al.[12] used Transmission Electron Microscopy (TEM) and Energy-Dispersive X-ray Spectroscopy (EDS) to report significant sulfur diffusion along grain boundaries when sulfur was deposited in the absence of oxygen. Taylor et al.[13] also used EDS and reported significant sulfur diffusion at high temperatures at the grain boundaries. Herndon et al. [10] and Li et al.[11] also reported sulfur diffusion at the grain boundary using Near-Field Optical Microscopy (NSOM) and Scanning TEM (STEM). Kranz et al.[14] used inverted substrate geometry CdTe cells to show sulfur diffusion via grain boundaries using a combination of atom probe tomography and secondary ion mass spectrometry. Rojsatien et al.[15] showed evidence of sulfur diffusion all the way to the back contact. However, none of these studies characterize the grain boundaries and identify which ones promote impurity diffusion. As grain boundaries are inevitable in polycrystalline materials, it is crucial to distinguish between the grain boundaries that assist diffusion and those that do not. This may enable us to engineer grain boundaries during the deposition process to better mitigate detrimental impurity diffusion.

Light absorption, reflection, and hence the device efficiency can also be affected by grain sizes and orientations in the CdTe absorber layer[16]. The correlation between the number of stacking faults and twins in its microstructure and efficiency is not well understood.[17]. Hence, there is a need for a comprehensive characterization of grain boundaries (in terms of misorientations and grain boundary plane normals)[18] and simultaneously, mapping the relative location of impurities, and grain size/orientation distribution in a functioning solar cell to understand the correlation between microstructure and device efficiency.

To characterize grain boundaries, researchers have extensively used electron microscopy techniques such as TEM, Scanning TEM, and Electron Back Scattered Diffraction (EBSD)[19]. These are unsuitable for 3D characterization because of the limited penetration depth of electrons. Techniques based on X-ray fluorescence[20] are not sensitive to crystallographic ordering and, hence, are unsuitable for mapping grain boundaries. Recent developments in Bragg coherent diffraction imaging (BCDI)[21] and dark field X-ray microscopy[22] offer nanoscale resolution to resolve grain boundaries but are not suitable for 3D characterization of a polycrystalline sample with many grains due to limited field of view or incomplete sampling, which leads to incomplete statistics. Three-dimensional X-ray diffraction (3DXRD) is a well-established tool to comprehensively study polycrystalline samples and characterize the position, size, strain, and orientation of individual grains[23]. This technique can also be used for the non-ambiguous determination of phases with similar lattice parameters.[24] Scanning 3DXRD[25] (S3DXRD) is a variation of 3DXRD where a point beam is used to raster scan the sample. At the expense of an increased data acquisition time, this extends the capabilities of 3DXRD by improving the spatial resolution to the sub-micrometer range, which is the typical scale to be investigated for polycrystalline solar cell materials. Hayashi *et al.*[25] first used scanning 3DXRD to obtain an orientation map of an iron specimen. Hektor *et al.*[26] studied the evolution of a tin whisker under high temperatures using S3DXRD. In both cases and other investigations [27–30] employing diverse modalities of 3DXRD, examining grains with dimensions below 1 micrometer (µm) has proven challenging.

With the EBS (Extremely Brilliant Source) upgrade of the ESRF (European synchrotron radiation facility) and our modified approach of indexing using prior information (see methods section), we could characterize and visualize grain boundaries of grains as small as 300 nm. We could also map the strain field inside the grains with a resolution of 100 nm. This allows us to identify grain boundaries associated with high strain levels.

## 2. Results

The CdTe solar cell device was fabricated using the conventional superstrate configuration method[31]. To ensure sufficient X-ray transmission and avoid any spot overlap due to a high number of grains, we used Focused Ion Beam (FIB) to extract a cylindrical pillar-shaped sample out of the full solar-cell stack. The height and width of the cylinder were 5.2 and 5 µm, respectively. Scanning 3DXRD (schematic Figure 1A) with a beam size of 100 nm and exposure time of 5 milliseconds was performed at the ID11 beamline of ESRF[32] on the CdTe p-type absorber layer [36], and raw data was analyzed as described in the methods section.

### 2.1 Statistics of grain sizes, orientations, and lattice parameters

With a misorientation threshold of 0.2° between connected grains, we find 113 grains in our 3D volume of the CdTe absorber layer. The grain-size distribution is provided in Figure 1F. The mean grain size (equivalent radius of a sphere) is 0.95 µm, which agrees well with the results from the SEM image attached in the supplementary information (Figure S1.A). Figure 1C and Figure 1D show horizontal and vertical slices of the reconstructed grain map. It can be seen from these visualizations that most of the grain boundaries in the cross-sections are twin boundaries. These are $\sum 3$ (sigma three) twins (GBs colored yellow). The Inverse Pole Figure (IPF) density function plot, Figure 1G, reveals that the sample has a moderate texture. A minimum near the [111] direction can be observed. This is consistent with the fact that [111] direction is also the $\sum 3$ twinning axis, hindering growth in this direction. Note that the grain statistics are too poor for a classical texture analysis. CdTe has a cubic lattice in the case of perfect stoichiometry. We calculated the three lattice parameters a, b, and c (representing grain averages) for all grains. Interestingly, the magnitudes of all three lattice parameters reduce continuously and synchronously with decreasing distance to the CdS layer (Figure 1H). The change in lattice parameters is too small to arise due to mechanical strain. Hence, this isotropic reduction indicates a compressive strain caused by chemical substitution close to the CdS layer.

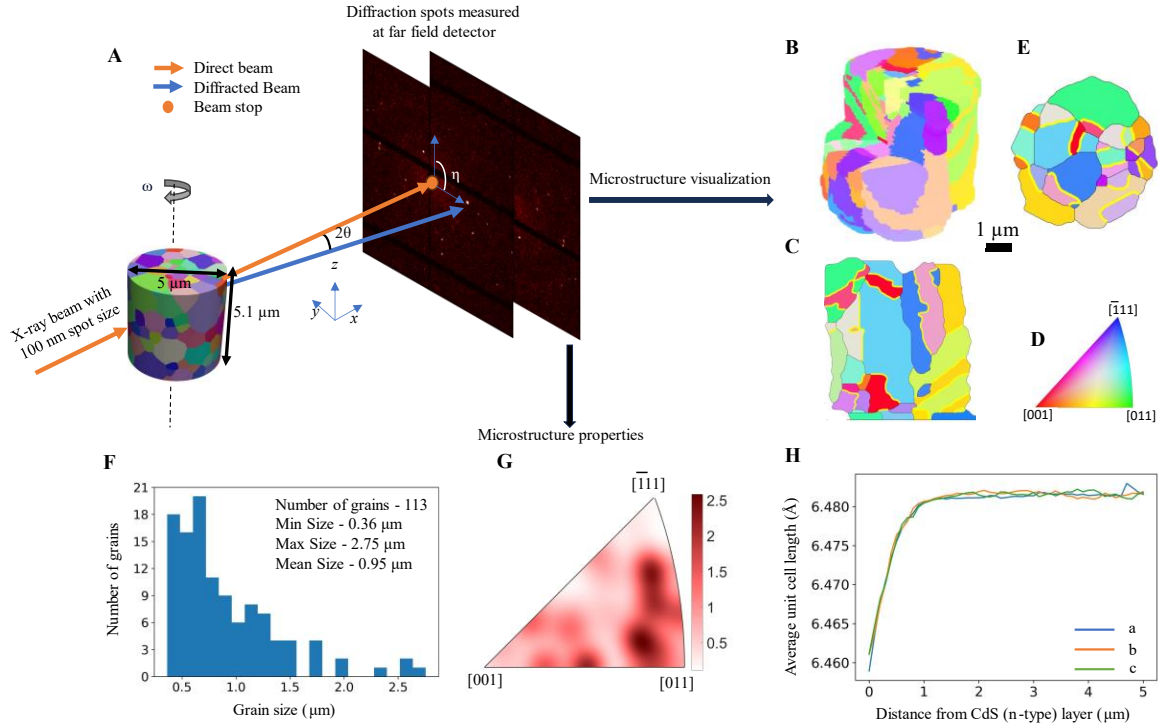

Figure 1. A) Schematic of scanning 3D-XRD setup. The laboratory coordinate system is defined. The rotation angle ω, the Bragg diffraction angle θ, and the azimuthal angle η characterize the diffracted beam for a reflection from some grain in the sample. B) 3D visualization of grain map of p-type CdTe solar cell. C, D) Orthogonal slices through the center of the sample. Yellow lines represent a twin grain boundary. E) Inverse Pole Figure (IPF) color scale for crystal orientation viewed along the Z-axis. F) Histogram depicting grain size distribution. G) Magnitude of Inverse pole figure density function. H) Plot showing the variation in average unit cell parameters a, b, and c with distance from the CdS n-type layer.

## 2.2 Grain shapes and grain positions

The aspect ratios of the grains (defined as the ratio of spherical grain size in the z-direction to the ratio of spherical grain size in the xy plane) are plotted as a function of grain size in Figure 2A. We observe a positive correlation. This means that the big grains tend to exhibit 'columnar' growth. This is consistent with the SEM image in the supplementary Figure S1A. The histogram in Figure 2B shows that 62.2% of large grains (grain size> mean grain size) grow from the CdS substrate.

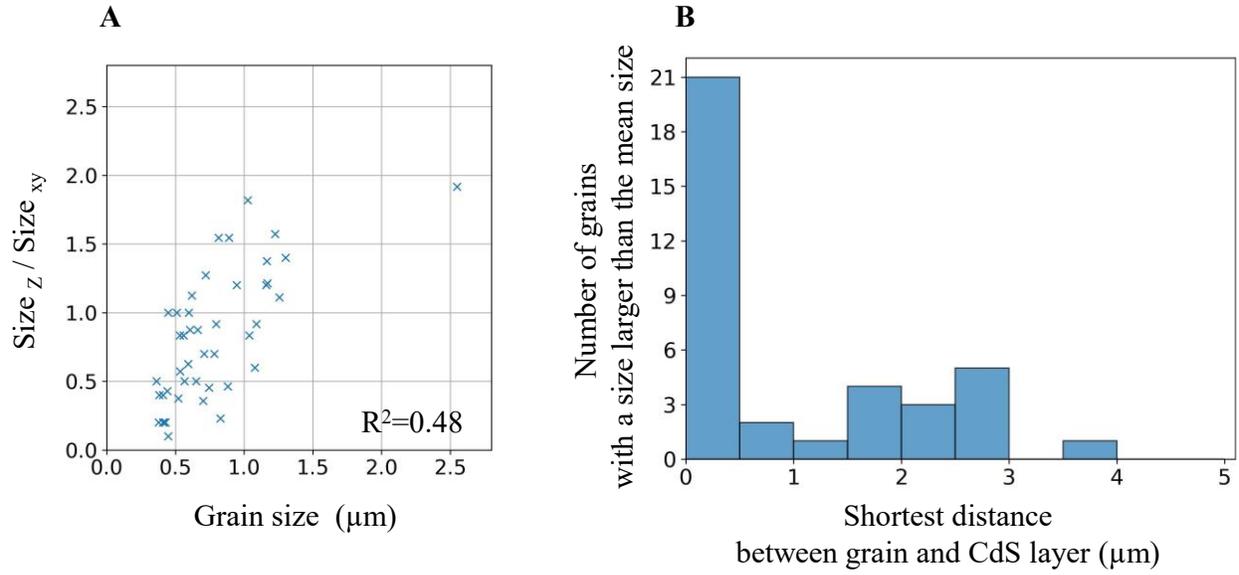

Figure 2. A) Plot showing the ratio between horizontal and vertical grain sizes against spherical grain sizes. B) Histogram of grains with a size larger than the mean size plotted as a function of their shortest distances to the CdS substrate

## 2.3 Distribution of twin domains

As seen in Figure 1C and Figure 1D, CdTe has many $\sum 3$ twin-domain boundaries. As discussed, these help relieve the tensile stress that forms during the growth phase of thin films. During the annealing of thin films under $CdCl_2$ environment (see Experimental section), the twin domains get thicker[6]. This leads to a decrease in the average grain size and increases the number of twin boundaries. The role of twin boundaries in assisting the diffusion of impurities and promoting charge carrier recombination is debated[33,34,35]. As the direction of grain boundaries is columnar (perpendicular to the substrate, see Figure 1D), twin boundaries tend to form parallel to the substrate. Figure 3A shows a group of 7 grains related by a twin relationship. After merging neighboring grains with a $\sum 3$ twin relationship, we find 30 twin families. Moreover, 88.7% of the sample volume can be attributed to a twin family growing from the substrate. Hence, the formation of twins leads to a significant increase in the number of grains and grain boundaries in the bulk of CdTe.

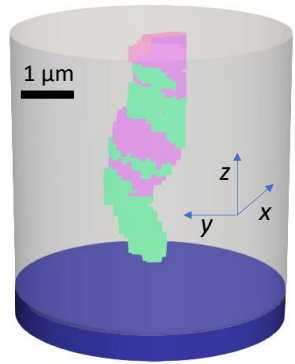

Figure 3. A) 3D visualization of a family of 7 twin-related grains from an appropriate viewing angle. The twins have only two orientations (represented by green and purple colors). 88.7% of the sample volume can be attributed to a twin family growing from the substrate. In this and the following figures, the transparent cylinder in the reconstruction illustrates the sample volume. The purple disk at the bottom shows schematically the n-type CdS layer.

**2.4 Correlations between strain and grain boundaries.**

As highlighted in the introduction section, it is well known that sulfur and other impurities diffuse into the bulk of CdTe and segregate along the grain boundaries. This is corroborated by our EDS-SEM spectra (see supplementary information, Figure S1.B), showing sulfur present in the bulk of the CdTe layer. It was reported that sulfur can replace tellurium in the lattice, causing a uniform contraction of the unit cell[11]. By using S3DXRD to measure the local strain with high accuracy we can therefore and hence can give an estimate of the concentration of impurity present.

We measured the intragranular strain tensor in the absorber layer using the procedure outlined in step 7 of the analysis pipeline (see experimental methods). As mentioned, an isotropic contraction of the unit cell was detected in the grains close to the CdS layer, as depicted in Figure 1H. Consequently, for our subsequent analysis, we exclusively considered the hydrostatic strain. We observe that strain depends strongly on distance from the CdS layer (Figure 1H). In the proximity of the CdS layer (~1 µm), we observe numerically high strain values (<-0.001) without any clear correlation with the position of grain boundaries and grain orientations. On average, the strain values continuously decrease with distance from the CdS layer. Interestingly, there are some areas in the 3D volume of the CdTe layer where high values of local strain are present within the bulk at large distances from the CdS layer (see supplementary information, Figure S2). Two of these areas are shown in Figure 4 and Figure 5.

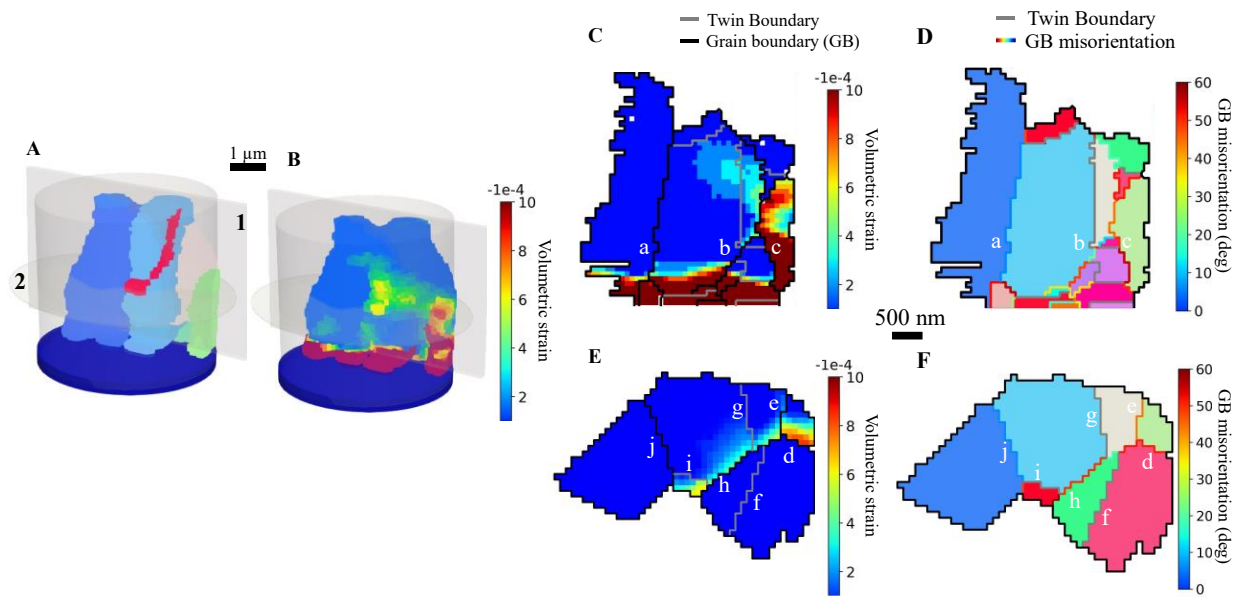

Figure 4. A) and B) are 3D visualizations of grain orientation and strain, respectively, from an appropriate viewing angle within the CdTe p-type layer for a selected region where grains show high strain and strain localization. In A, the color scale is the IPF color code for cubic symmetry. In B, the color scale is saturated near the CdS layer to emphasize weak strain in the bulk. C) Vertical cross-section of B) along plane 1 showing strain with annotation of boundary types. D) Grain orientations in the same plane as C) with annotation of grain boundary character. E) and F) Similar cross-sections now along the horizontal plane 2 in A. Grain boundaries are marked with lowercase letters. Neighboring grains are included in Figures C), D), E), and F), which form the strained grain boundary.

It is evident from Figure 4 that strain along grain boundaries marked c, d, and h in Figure 4C and Figure 4E, respectively, is significantly higher than along other grain boundaries and in the surrounding region. Table 1 shows the average misorientation angles along various grain boundary(s), grain boundary types, and if the grain boundary shows strain localization.

| Grain Boundary (s) | Average misorientation (deg) | Is a twin boundary? | Strain localization |
|---|---|---|---|
| a | 13 | No | No |
| b | 35 | Yes | No |
| c | 54 | No | Yes |
| d | 55 | No | Yes |
| e | 43 | No | No |
| f | 60 | Yes | No |
| g | 60 | Yes | No |
| h | 55 | No | Yes |
| i | 60 | Yes | No |
| j | 14 | No | No |

Table 1: Properties of grain boundaries in the region shown in Figure 4.

It can be inferred from Table 1 that grain boundaries with high misorientations are preferred as sites for strain localization. It is also important to note that twin boundaries do not show strain localization. Figure 5 displays results for another region in the 3D volume, where high strain values are observed in the bulk. Figure 5B shows strain homogeneously spread over two different grains near the CdS layer. Again, as seen before, strain decreases rapidly as a function of distance from the CdS layer. Figure 5C shows a vertical cross-section of Figure 5B. As was seen in Figure 4, strain localization takes place along grain boundaries with high misorientation and does not take place along twin boundaries. Supplementary table S3 compares various grain boundaries of this region, their misorientation, and strain.

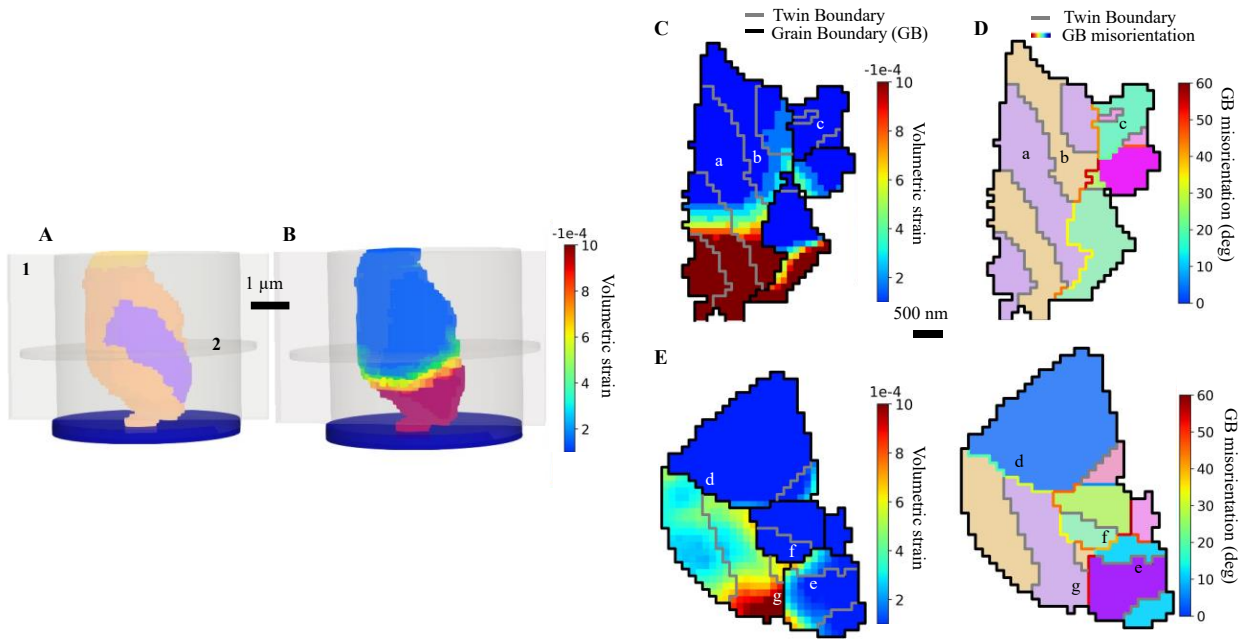

Figure 5. A) and B) 3D visualizations of grain orientation and strain, respectively, from an appropriate viewing angle, for a selected region where grains show high strain and strain localization. In A, the color scale is the IPF color code for cubic symmetry. In B, the color scale is saturated near the CdS layer to emphasize weak strain in the bulk C) Vertical cross-section of B) along plane 1 showing strain with annotation of boundary types. D) Grain orientations in the same plane as C) with annotation of grain boundary character. E) and F) Similar cross-sections now along the horizontal plane 2 in A. Grain boundaries are marked with lowercase letters. Neighboring grains are included in Figures C), D), E), and F), which form the strained grain boundary. We observe that high misorientation angle grain boundaries b, f, and g allow localization of strain, whereas twin boundaries colored in grey do not show localization of strain.

## 3. Discussion

The 3-dimensional reconstruction of the grains and the strain fields therein (Figures 4 and 5) reveal that close to the CdS layer, there are high strain values (near grain boundaries and in the grain interior). Further away from the CdS layer, the strain field is localized near high misorientation grain boundaries. Twin boundaries do not show any strain within the resolution of the insturment. Additionally, the strained grain boundaries form connected regions in space extending from the CdS layer (see Supplementary Figure S2). The observation that the observed compressive strain is isotropic (Figure 1E), is compatible with the attribution of the strain along these connected regions in space to the diffusion of sulfur from the CdS layer into the CdTe layer. This diffusion mechanism involves the substitution of tellurium with sulfur in the CdTe lattice, resulting in unit cell contraction. As sulfur diffuses inside the bulk of CdTe, it leaves a footprint of contracted unit cells along its pathway, resulting in this observed connected region of high strain. This is in agreement with a decreased Te/Cd ratio in the bulk towards the CdS interface as confirmed by X-ray fluorescence data (See supplementary Figure S4) that were obtained as complementary modality simultaneously as the S3DXRD data. From our results above, we infer that the diffusion pathways are highly selective. In the first stage, close to the CdS layer (< 1 μm from the CdS layer), sulfur diffuses homogeneously into both grain boundaries and the bulk of the grains. In the next stage (> 1 μm from the CdS layer), sulfur diffusion reduces significantly and only occurs along grain boundaries with a high misorientation. There is no observable diffusion along the twin boundaries. This can be because the density of dangling bonds and, hence, free energy is higher at high misorientation grain boundaries but very small at the twin boundaries.

**Outlook**

Scanning 3DXRD is an optimal characterization tool to simultaneously study grain orientations and strain with a high spatial resolution. With the recent developments in instrumentation and reconstruction algorithms, it is possible to reach a spatial resolution of 100 nm and strain resolution of $10^{-4}$. This enables us to image grain boundaries and, at the same time, quantify the 3D strain field present around these grain boundaries. We believe that studying the diffusion of impurities in solar cells, movement of ions in batteries, and shape changes in piezoelectric/ferroelectric materials can be made simpler using S3DXRD. Characterizing different grain boundaries and their environment facilitates grain boundary engineering. This new path to material design is critical to all polycrystalline solar cell technologies, which face the challenge of high non-radiative recombination at grain boundaries.

**Conclusions**

Scanning 3DXRD was used to characterize grain boundaries of a highly twinned CdTe solar cell absorber layer in 3D, visualizing grains as small as 300 nm with a resolution of 100 nm. It was observed that grain sizes are reduced, and grain orientations parallel to the growth direction are affected by twinning in CdTe. Our analysis of grain size distribution and orientation also reveals that grains tend to show columnar growth and most of the grain orientations arise due to twin formation of grains growing from the substrate. This grain growth model in CdTe can help researchers develop ways to deposit grains with favorable sizes and orientations. Using the 3D visualizations of grain orientations and strain, we observe that strain localization occurs near high misorientation grain boundaries and does not occur along twin boundaries. We suggest that the observed strain is due to sulfur diffusion from the CdS substrate into the CdTe lattice. Hence, this work enables us to modify and improve our understanding of sulfur diffusion into the bulk of CdTe/CdS devices, allowing us to optimize our deposition processes for CdTe solar cell devices.

**Methods**

A. Sample preparation

The CdTe solar cell device was fabricated using the conventional superstrate configuration method[31] for CdTe solar cells. The device is illustrated in Figure 6. Device fabrication starts with a commercial transparent conducting oxide (TCO) coated soda-lime glass. It is a bilayer stack with 500 nm of conducting $SnO_2$: F as the bottom layer and a 100 nm insulating layer of undoped $SnO_2$ on top. On top of the TCO, an 80 nm thin film of CdS is sputtered as the n-type buffer layer. Next, the p-type CdTe absorber layer is deposited using close-space sublimation to achieve a thickness of approximately 5 μm. At this stage, the device was heat-treated with $CdCl_2$ at 420° C. As a next step, back contact is deposited using evaporation of 2.5 nm of Cu, followed by sputtering deposition of 375 nm of ZnTe and 500 nm of Mo. The stack is then annealed at 230° C.

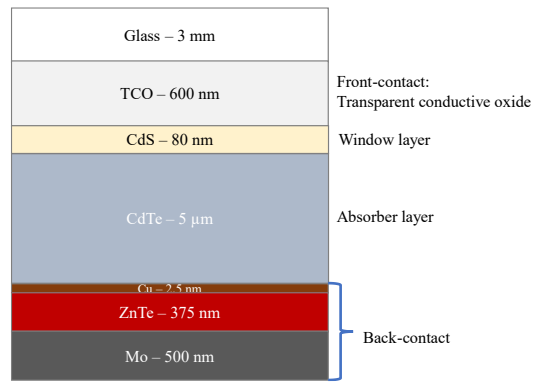

Figure 6. Schematic of CdTe solar cell

This device manufacturing process is consistent with many manufacturing practices in the CdTe solar cell industry. The device shows an open-circuit voltage of 850 mV, 22 mA/cm$^2$ short-circuit current density, and 75% fill factor, resulting in 14% conversion efficiency under standard measurement conditions (1000 W/m$^2$ intensity, AM1.5G spectrum).

B. Sample Modification for S3D-XRD

In 3DXRD experiments, it is crucial to ensure that the diffraction spots are well separated on the detector. This means an ideal sample should be small enough to minimize the measurement time and avoid too many grains being under illuminated at the same time. We used the Focused Ion Beam (FIB) lift-out technique to achieve this. In this technique, the desired specimen is extracted from the bulk sample and is mounted on a suitable sample holder using platinum micro-welding. The entire procedure is performed inside an electron microscope. Hence, we transformed our solar cell layer stack into a cylindrical pillar ~5 μm in diameter and ~5.2 μm in height (From CdS top layer to Molybdenum, bottom layer).

C. Experimental Setup

The nasoscope station at the ID11 beamline of the European Synchrotron Radiation Facility (ESRF) in Grenoble, France, was utilized for the synchrotron X-ray experiment. A monochromatic beam with an energy of 42 keV was focused to a spot size of 100 nm using a pair of silicon compound refractive lenses (CRL). Fluorescence measurements were done simultaneously to map the sample boundary and account for sample drift accurately. To cover the sample volume using the 100 nm beam, we raster scanned the sample in steps of 100 nm along y and z (refer to the coordinate system in Figure 1A). The area of the raster scan was 8 μm * 5.5 μm respectively. Diffraction patterns at each scan point were captured using an Eiger2 4M CdTe detector situated 144 mm behind the sample, with a pixel size of 75 μm and an exposure time of 0.005s. The step size for rotation was 0.125°. An illustration of the setup and reference to the coordinate system can be found in Figure 1A.

D. Data Reduction and Analysis

Every layer in z is analyzed independently. To generate the grain orientation maps and strain maps, the 2D diffraction patterns were subjected to a series of processing steps-

1. *Identification of diffraction spots* –The 2D diffraction patterns are converted to sparse frames to handle the large amount of data. After detector distortion correction and background subtraction, a suitable threshold is applied to identify diffraction spots and extract their center of mass positions and integrated intensities.
2. *Merging of diffraction spots* – Spots identified in 2D patterns are merged if the center of mass detector positions are close for the neighboring frames in y and w steps.
3. *Calibration* –We assign scattering vectors $\vec{G}$ to every diffraction spot based on the center of mass detector positions and the experimental geometry.
4. *Indexing* – The $\vec{G}$ vectors are then assigned to grains, and at the same time, average orientation **U** and reciprocal space metric **B** are found for each grain. The Laue condition for diffraction implies
$$\mathbf{UBI}\vec{G} = \vec{G}_{hkl} \qquad (1)$$
Here $\vec{G}_{hkl}$ is the scattering vector in reciprocal space, and **UBI** is the inverse of **UB**. Using the *ImageD11* software[37], we find the UB matrices such that the left-hand side of Eq (1) gives integral values of h, k, and l within a tolerance value.
5. *Grain shapes* – After finding the orientation and peaks for every grain, sinograms are constructed for each grain by using diffraction intensities from the grain at various scan points and rotation angles. Filtered back projection is used to obtain a grain shape.
6. *Stitching of grains in 2D* – All grain shapes are projected to a grid of 2D voxels (for a particular value of z). Grain having more intensity post-normalization is assigned to the voxel. As grain density is lower at the grain boundary interface, grain boundaries can be located using the contrast in intensities.
7. Steps 1-6 are repeated for every z position in the dataset.
8. *Voxel (Intragranular) orientation and strains* – This involves refining the peak list for each grain using the obtained voxel positions and fitting the **UBI** matrix to the new peak list for the grain. Intragranular orientation and strain are obtained by the decomposition of the **UBI** matrix into **U** and **B**.

All these processing steps have been covered in detail before [24,26,36–38].

E. Finding small grains in bulk

Grain mapping of nanocrystalline grains has been challenging because small grains typically have low diffracted intensity. This makes it harder to separate diffraction spots from small grains and weak spots arising from neighboring positions in the sample illuminated unintentionally by the beam tail. We solved this problem by adding another tolerance factor based on misorientation to the indexing step to remove any noise orientations arising from beam tails (see flowchart 1). This ensures that a particular grain orientation is found in more than one layer in the z-direction of the raster scan.

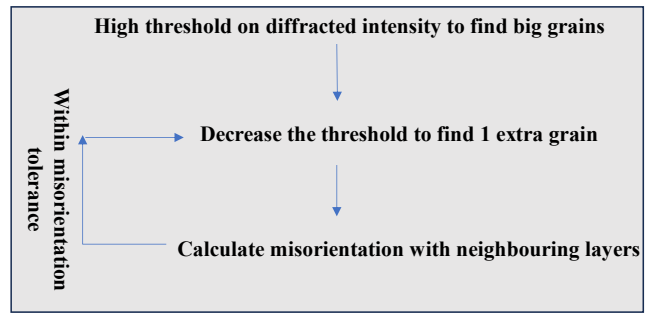

Figure 7. Using information from neighbouring layers for indexing

Misorientation analysis is also helpful in cases with twinned grains. Twinned grains can share (60-70%) peaks between grains. This makes stitching grains in 2D (step 6 in the reconstruction process outlined before) give non-unique solutions. Adding a misorientation tolerance factor with the orientation above or below the voxel helps in this step. This method can be extended to letterbox beam 3DXRD[24] to make more accurate center-of-mass maps for each layer in z. (Grain size bigger than 5 μm).

**Acknowledgments**

Contract No. DE-AC36-08GO28308. Funding provided by U.S. Department of Energy Office of Energy Efficiency and Renewable Energy Solar Energy Technologies Office. The views expressed in the article do not necessarily represent the views of the DOE or the U.S. Government. The U.S. Government retains and the publisher, by accepting the article for publication, acknowledges that the U.S. Government retains a nonexclusive, paid-up, irrevocable, worldwide license to publish or reproduce the published form of this work, or allow others to do so, for the U.S. Government purposes. These experiments were carried out with the help of scientists and people responsible for ID11 ESRF and P03 nano focus at PETRA III. We acknowledge support from the Danish ESS lighthouse on hard materials in 3D, SOLID. Special thanks to the DANSCATT instrument center, funded by the Danish Ministry for Higher Education and Research, for supporting the travel expenses during the experiment.